\documentclass[twocolumn]{aastex63}
\usepackage{multirow}
\usepackage{natbib}
\usepackage{lineno}

\def\h{\hskip 0.0 mm}

\def\apjl{{ApJL}}
\def\apjs{{ApJS}}
\def\asec{$^{\prime\prime}$}

\def\e#1{$\times$10$^{#1}$}

\def\farcs{\hbox{$.\mkern-4mu^{\prime\prime}$}}

\def\hal{H$\alpha$}
\def\hb{H$\beta$}

\def\lax{{$\mathrel{\hbox{\rlap{\hbox{\lower4pt\hbox{$\sim$}}}\hbox{$<$}}}$}}
\def\gax{{$\mathrel{\hbox{\rlap{\hbox{\lower4pt\hbox{$\sim$}}}\hbox{$>$}}}$}}
\def\simlt{\lower.5ex\hbox{$\; \buildrel < \over \sim \;$}}
\def\simgt{\lower.5ex\hbox{$\; \buildrel > \over \sim \;$}}
\def\lum{erg s$^{-1}$}
\def\mbh{{$M_{\rm BH}$}}

\def\cm2{cm$^{-2}$}

\def\oiii{[\ion{O}{3}]}

\def\lbol{$L_{{\rm bol}}$}

\def\mm{$M_{\rm BH}-M_{\star}$}

\def\msig{$M_{\rm BH}-\sigma_\star$}

\shorttitle{MIR Variability of low$-z$ AGNs}
\shortauthors{KIM et al.}

\begin{document}

\title{Mid-infrared Variability of Low-redshift Active Galactic Nuclei : 
Constraints on a Hot Dust Component with a Variable Covering Factor}

\author{Suyeon Son}
\affiliation{Department of Astronomy and Atmospheric Sciences,
Kyungpook National University, Daegu 702-701, Korea}

\author{Minjin Kim}

\affiliation{Department of Astronomy and Atmospheric Sciences, 
Kyungpook National University, Daegu 702-701, Korea}

\author{Luis C. Ho}

\affiliation{Kavli Institute for Astronomy and Astrophysics, Peking 
University, Beijing 100871, China; lho.pku@gmail.com}

\affiliation{Department of Astronomy, School of Physics, Peking University, Beijing 100871, China}

\correspondingauthor{Minjin Kim}
\email{mkim.astro@gmail.com}

\begin{abstract}
We utilize mid-infrared multi-epoch data from Wide-field Infrared Survey Explorer (WISE) 
over a $\sim10$-year period in W1 (3.4$\mu$m) and W2 (4.6$\mu$m) bands to investigate 
the structure of dusty torus in low-redshift ($0.15<z\le 0.4$) active galactic nuclei (AGN). 
We calculate a Spearman correlation coefficient ($r_{12}$) 
between W1 magnitude and W1$-$W2 color based on the light curve in individual
objects. Interestingly, $r_{12}$ spans a broad range from $-1$ to $1$,
and is detected to be correlated with mean W1$-$W2 color and 
AGN bolometric luminosity, in the sense that objects with blue W1$-$W2 color
and low AGN luminosity tend to become redder (bluer) with increasing 
(decreasing) W1 brightness in the light curve (i.e. $r_{12} < 0$),
although the correlation of $r_{12}$ with the bolometric luminosity is 
relatively weak.
The fit for the spectral energy distribution reveals a significant contribution 
from the host galaxy in the W1 and W2 bands. However, the dependencies of $r_{12}$
on W1$-$W2 color and AGN luminosity still persist even after careful 
elimination of the host light contribution. We propose that this result 
can be explained if covering factor of the hot dust component decreases as 
AGN luminosity increases.     
\end{abstract}


\section{Introduction} 
An accretion disk in the vicinity of a supermassive black hole and dense gas clouds
known as the broad line region (BLR), is assumed to make up the central part of 
the active galactic nucleus (AGN), which is surrounded by a dusty torus. According to
the unification model, the structures of type 1 and type 2 AGN are identical
but the AGN type is solely determined by a viewing angle to the 
opening angle of the dusty torus \citep{antonucci_1993, urry_1995}. In this 
regard, it is critical to understand the detailed structure of the torus 
to examine the AGN unification model.  

While clumpy dust instead of smooth dust may mimic the geometrical structure 
of the torus  \citep[e.g.,][]{pier_1992, fritz_2006, nenkova_2008, honig_2010, 
stalevski_2016}, 
several critical concerns about the structure of the torus structure
remain unresolved. For example, the covering factor of the torus is one
of the most essential parameters to determine the torus structure. Various 
observational studies reached somewhat different conclusions. Some studies 
claimed that the covering factor is somewhat linked with Eddington ratio 
\citep[e.g.,][]{ricci_2017, ezhikode_2017, zhuang_2018} and independent of AGN 
luminosity \citep[e.g.,][]{hao_2010,hao_2011,mor_2012, netzer_2016}. 
However, using the ratio of mid-infrared luminosity and AGN bolometric 
luminosity, various other studies 
demonstrated that the covering factor is anti-correlated with AGN luminosity
\citep[e.g.,][]{maiolino_2007, mor_2011, lusso_2013, roseboom_2013}, favoring 
a receding torus model \citep{lawrence_1991, simpson_2005, honig_2007}.

\begin{figure*}[thp]
\centering
\includegraphics[width=0.95\textwidth]{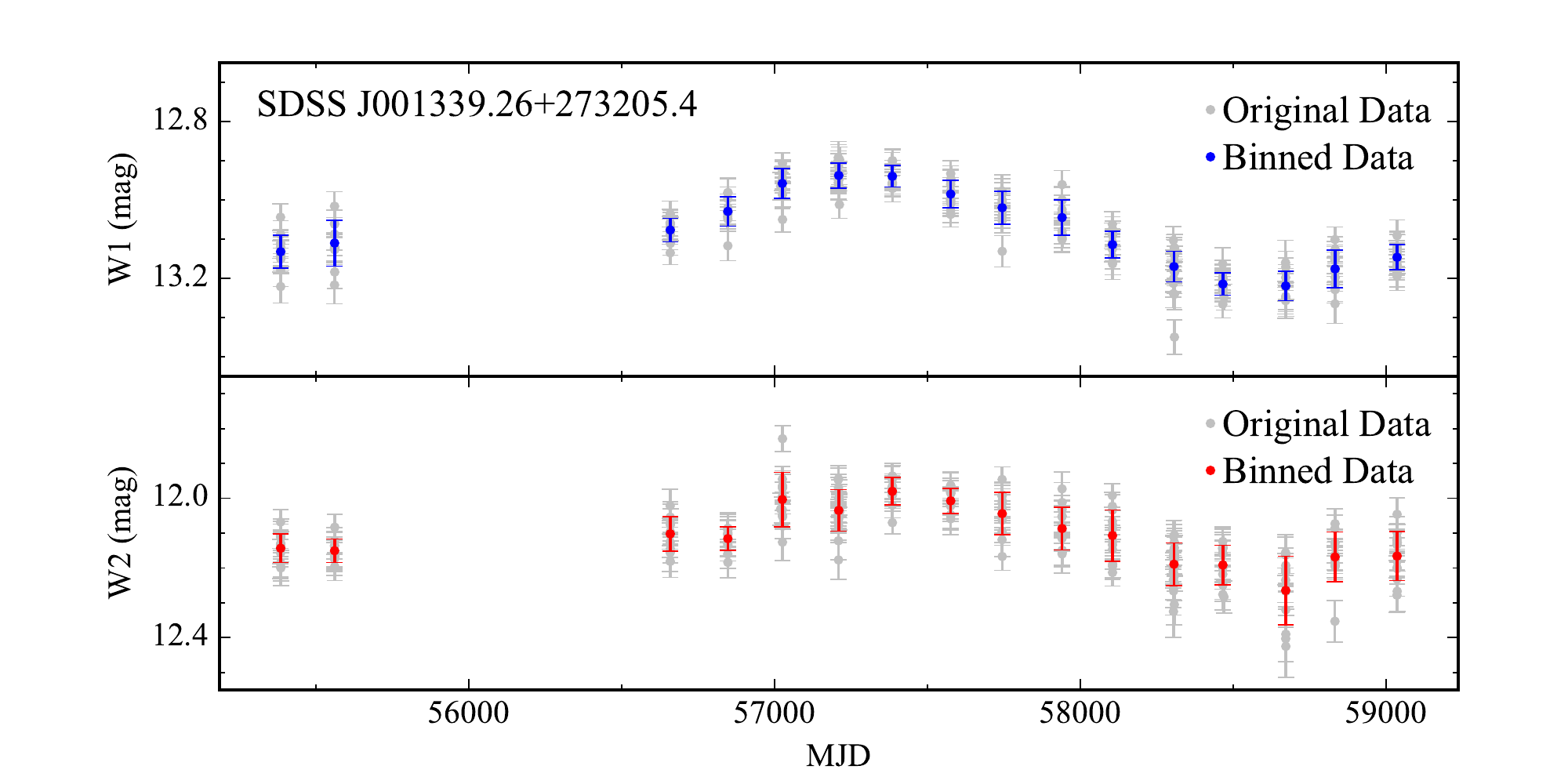}
\caption{
An example of light curve. Blue and red represent the light curves   
after binning for W1 and W2 bands, respectively,
while grey denotes the original measurements before binning. 
}
\end{figure*}

Spectral energy distribution (SED) of near-infrared (NIR) and mid-infrared 
(MIR) is critical for constraining the physical properties of the dusty torus
as the dust heated by the immense energy from the accretion disk re-radiates
thermal emission in the infrared (IR). It is well known that the IR SED can 
be represented by hot and warm components \citep[e.g.,][]{barvainis_1987,
suganuma_2006,mor_2012}, although an additional contribution from a polar dust 
may not be insignificant \citep[e.g.,][]{honig_2013, lyu_2018, yang_2020,
mountrichas_2020, buat_2021, toba_2021}. 
Several observational studies discovered a population 
of AGN lacking the hot dust component at moderate and high redshift, 
based on the IR SED of AGN
\citep[e.g.,][]{jiang_2010, hao_2010, hao_2011}. In addition, \citet{hao_2011} 
claimed that a fraction of hot-dust-deficient (HDD) AGN in the distant Universe
is greater than that in the local universe. However, other studies 
demonstrated that a population of HDD AGN may not be correlated with the redshift 
but firmly connected to the physical properties of AGN, such as 
bolometric luminosity and Eddington ratio \citep[e.g.,][]{mor_2012, jun_2013, 
lyu_2017}. 

Despite various previous studies, it is still vital to understand the 
structure of the dusty torus and how it relates to the physical properties 
of AGN. In this study, we utilize an independent approach based on the 
characteristics of MIR variability to investigate the physical properties of 
the dusty torus. The sample selection and multi-epoch data are presented 
in Section 2. We describe the 
approach to quantify the characteristics of MIR variability in Section 3. In 
section 4, we discuss the physical origin of our finding. Throughout the study,
we assume a cosmology with $H_0=70$ km s$^{-1}$ Mpc$^{-1}$, 
$\Omega_m=0.3$, and $\Omega_{\lambda}=0.7$.   

\section{Sample and Data}

\subsection{Sample}
Our sample is drawn from the Sloan Digital Sky Survey (SDSS). We 
select type 1 and type 2 AGN from the SDSS Data Release 14 (DR14) quasar 
catalog \citep{paris_2018} and the SDSS Max Planck Institute for 
Astrophysics-Johns Hopkins University (MPA-JHU) Data Release 8 (DR8) catalog 
\citep{kauffmann_2003, brinchmann_2004}, respectively. 
Because the MPA-JHU catalog contains all types of galaxies, we 
utilized only galaxies classified as AGN according to the 
Baldwin-Phillips-Terlevich (BPT) diagram \citep{baldwin_1981} in MPA-JHU 
classifications (BPTCLASS = 4; \citealt{brinchmann_2004}). 
For both types of AGN, we impose a minimum redshift of 0.15 to minimize 
the host galaxy contamination from extended features in  W1 (3.4 $\mu$m)
and W2 (4.6 $\mu$m) bands,
and a maximum redshift of 0.4 to ensure that \hal\ is covered by the SDSS spectrum 
for the AGN classification using the BPT diagram. For type 2 AGN, 402 
duplicates were removed in the MPA-JHU catalog. We find that 922 type 2 AGN
initially selected from the MPA-JHU catalog are also classified as type 1
in the SDSS QSO catalog, which are regarded as type 1 in this study, yielding
an initial sample of 7443 type 1 and 5531 type 2 AGN.

To investigate mid-infrared (MIR) variability of SDSS-selected AGN, 
we employed photometric data of W1 and W2 bands from 
the Wide-field Infrared Survey Explorer (WISE; \citealt{wright_2010}).  
Multi-epoch data are retrieved from the AllWISE Multiepoch Photometry Table 
and the NEOWISE-R Single Exposure (L1b) Source Table at the NASA/IPAC Infrared 
Science Archive (IRSA; https://irsa.ipac.caltech.edu/) with a matching radius of 
2\asec. Sources with detection in W1 band typically have an astrometric 
uncertainties less than 0\farcs5 \citep{cutri_2013}, 
indicating the matching radius is sufficient to capture 
the IR counterpart (see also \citealt{assef_2013}). 
Among the initial sample, 7368 type 1 and 5491 type 2 AGN were 
detected by WISE or NEOWISE. WISE visited a particular field in the sky 
every six months, and typically observed that field 14 times for three days in 
each visit. As a result, we bin the photometric data into $\sim\pm15$ days
to obtain a representative magnitude in each visit. 
To remove poor-quality data, 
only the photometry flagged with 
$cc\_flags = 0$, $qual\_frame > 0$, $qi\_fact > 0$, 
$saa\_sep > 0$, and $moon\_masked = 0$ were averaged after 3 sigma clipping 
in a single visit (Fig. 1). Following \citet{lyu_2019}, we evaluated the 
magnitude uncertainty for each epoch as follows:
$\sigma_\mathrm{epoch}^{2} = \frac{1}{N-1}\sum_{i=1}^N {(m_{i}-
\overline{m_\mathrm{epoch}})^2} + \frac{1}{N^2}\sum_{i=1}^N 
\sigma_{i,\mathrm{pho}}^2 + \frac{1}{N}\sigma_\mathrm{s.s.}^2$
where $m_{i}$ and $\sigma_{i,\mathrm{pho}}$ denote 
the magnitude and its uncertainty at each observation, respectively
in the Vega magnitude system, 
$\overline{m_\mathrm{epoch}}$ denotes the mean magnitude at each epoch,
and $\sigma_\mathrm{s.s.}$ denotes the system stability 
($\sim0.029$ and $\sim0.016$ mag for WISE and NEOWISE, respectively).
The light curve typically spans a baseline of 10 yr with a cadence of 6 
months. Note that there is a $\sim40$-month gap between WISE and NEOWISE.

\begin{figure}[t]
\centering
\includegraphics[width=0.45\textwidth]{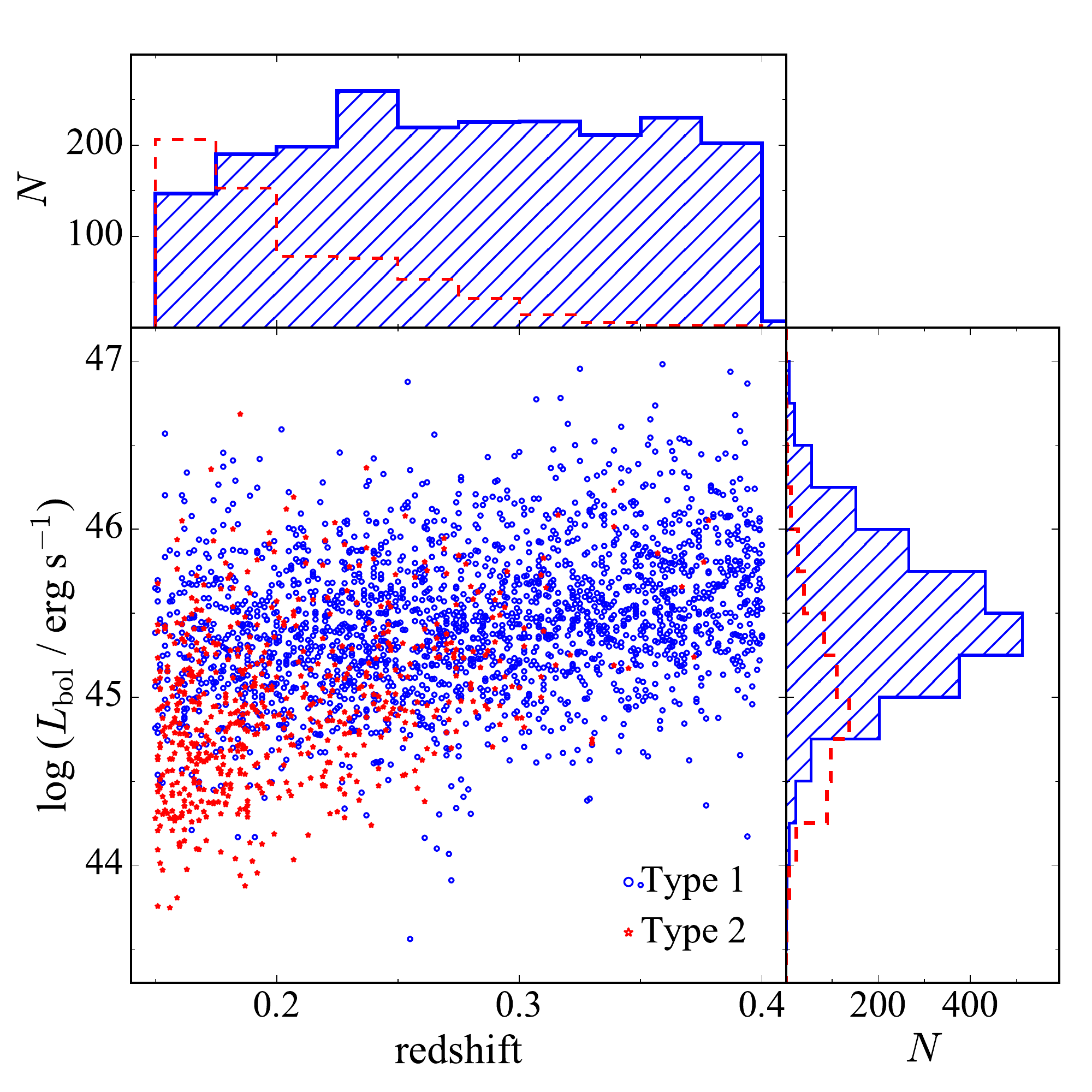}
\caption{
Redshift and bolometric luminosity distributions of the sample. type~1 AGN 
are denoted by blue stars and blue histograms, while type~2 objects are given 
by red circles and red histograms. Bolometric luminosity is estimated from the 
\oiii\ luminosity with a bolometric correction of 3500. 
}
\end{figure}

\subsection{MIR Variability}
To identify variable objects, we utilize the $\chi^2$ of the light curve: 
$\chi^2=\sum_{i=1}^{N}{[\frac{m_i-\bar{m}}{\sigma_i}]^2}$, where $\bar{m}$
represents the mean magnitude over the entire epochs and $\sigma_i$ is the uncertainty 
in each epoch. Following the recipe in \citet{sanchez_2017},
we defined $P_{var} \equiv 1-P(\chi^2)$, where $P(\chi^2)$ is the probability
that the computed $\chi^2$ or lower value can be obtained by chance from an 
invariable source (see also \citealt{mcLaughlin_1996}). 
Therefore a low $P(\chi^2)$ (high $P_{var}$) implies that the object is 
likely to be variable. We classified it as a variable object if $P_{var}>0.95$ 
both for W1 and W2 \citep{lanzuisi_2014,cartier_2015, sanchez_2017, kim_2018}.
Note that we utilized a sample with a sufficiently large number of epochs 
($N_\mathrm{epoch}\geq10$).
Due to larger uncertainty in W2 compared to W1 as shown in Figure 1,
the main reason for the rejection is relatively smaller $P_{var}$ in W2 band.
Finally, a total of 2114 type 1 AGN and 
623 type 2 AGN were identified as variable sources. We conducted further 
investigation for these sources. 

\begin{figure}[t!]
\centering
\includegraphics[width=0.45\textwidth]{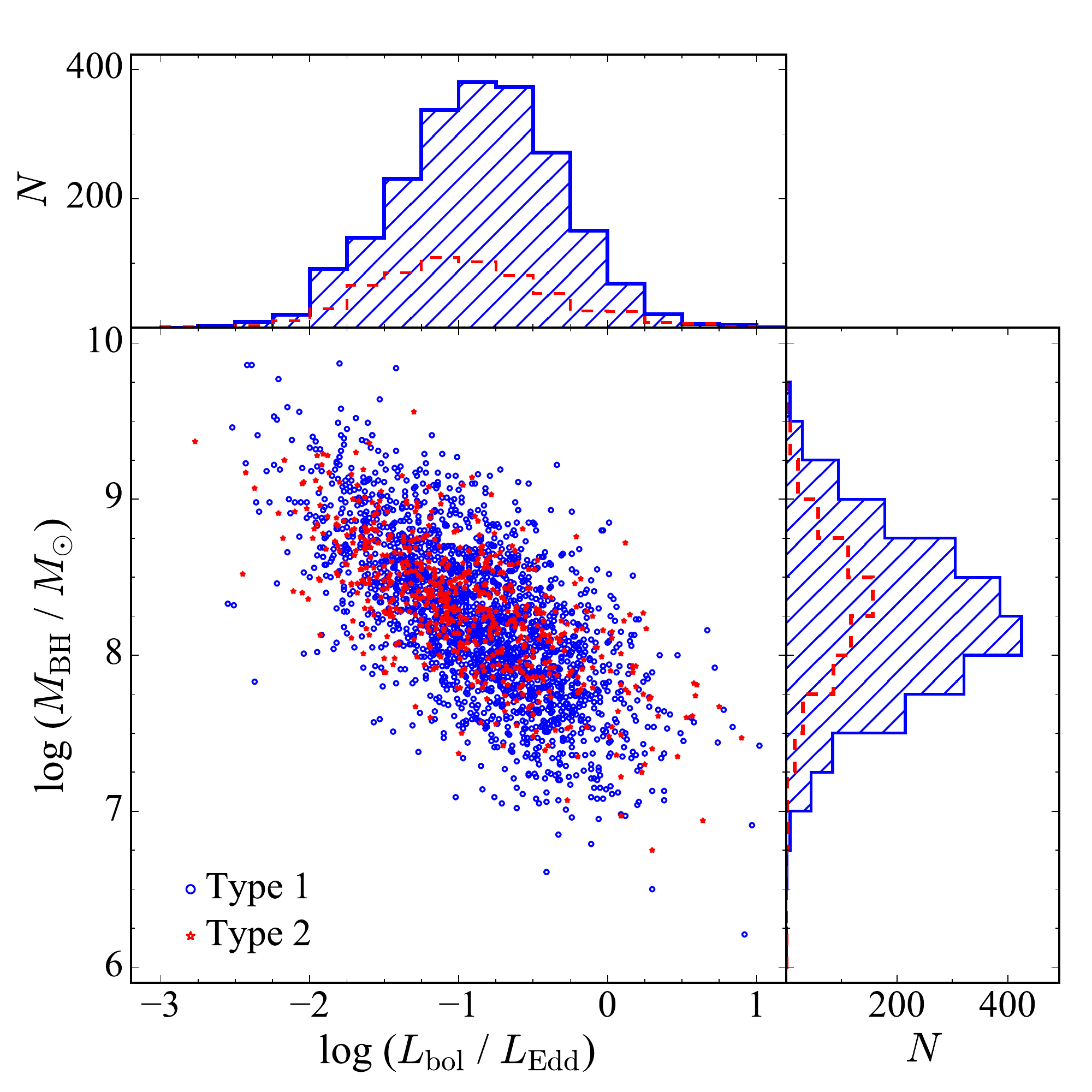}
\caption{
The distributions of BH mass and Eddington ratio of the sample. \mbh\ is 
measured using the virial method and the \mm\ relation for the type~1 AGN
and type~2 AGN, respectively. 
}
\end{figure}

\subsection{AGN properties}
We estimated the bolometric luminosity $L_\mathrm{bol}$ using 
\oiii$\lambda$5007 luminosity by adopting a conversion of 
$L_\mathrm{bol}/L_\mathrm{O\,III} \approx 3500$ \citep{heckman_2004}.
%which is suitable for \oiii\ luminosity not corrected for the extinction. 
Note that the conversion factor can be significantly smaller or dependent 
upon the AGN luminosity for the extinction corrected for \oiii\ luminosity
\citep{kauffmann_2009, trump_2015, kong_2018}. 
However, because our goal is to examine a trend of MIR variability
as a function of $L_\mathrm{bol}$, adopting a single universal 
bolometric correction from \citet{heckman_2004} is sufficient for that purpose.
\oiii\ luminosity of type 1 and type 2 AGN were taken from \citet{rakshit_2020}
and MPA-JHU DR8 catalog, respectively.
The distributions of bolometric luminosity and redshift of the sample
are shown in Figure 2. Note that there is a clear difference between 
type 1 and type 2 subsamples, in the sense that type 1 AGN have greater 
bolometric luminosity and larger redshift compared to type 2 AGN. 
Despite of this genuine difference, two samples show similar trends in 
the characteristics of MIR variability as shown in Section 3. It reveals 
that the main result of this study do not dependent on this bias.

\begin{figure*}[t!]
\centering
\includegraphics[width=0.95\textwidth]{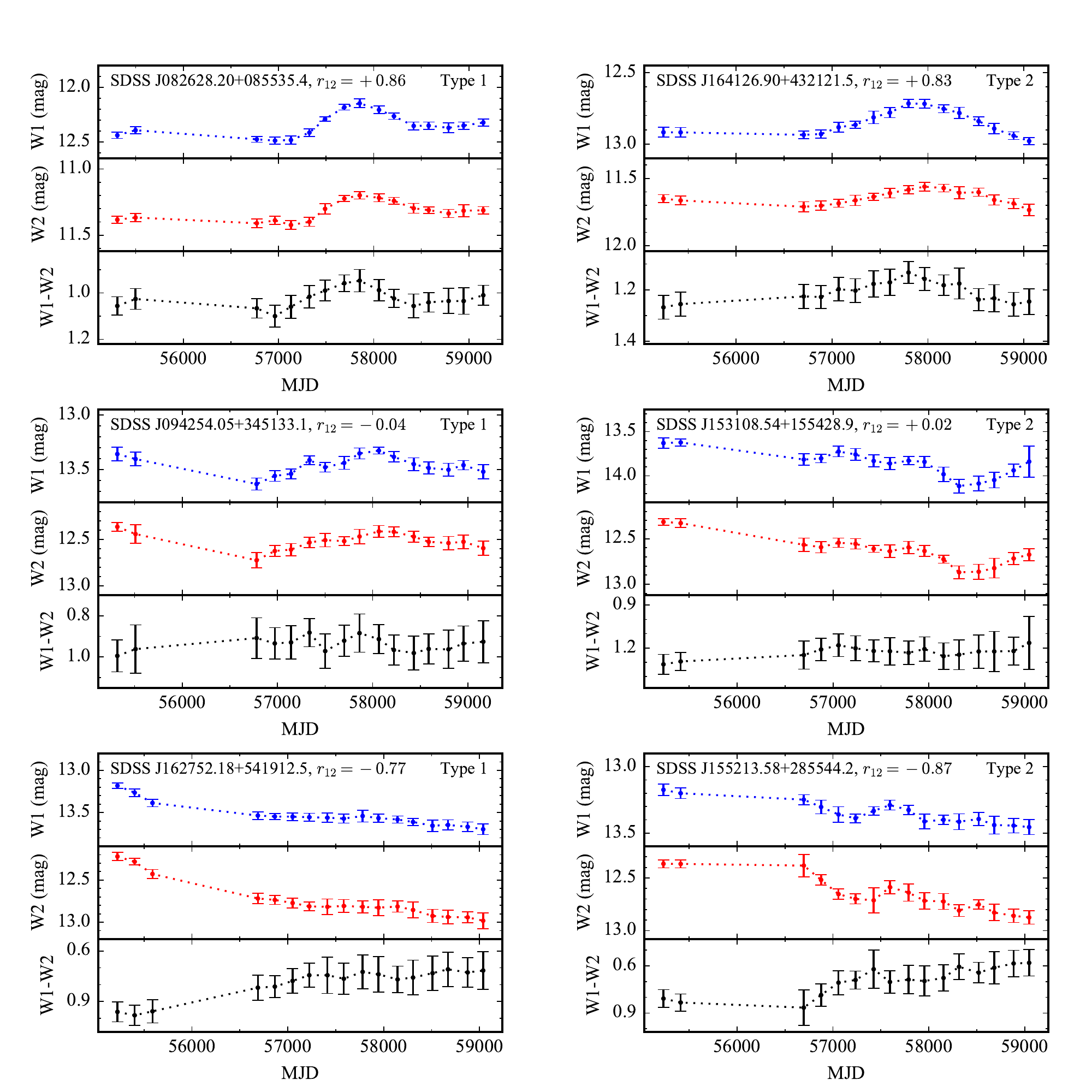}
\caption{
{Examples of light curves with strong correlations with a positive 
$r_{12}$ (top), weak correlations with $r_{12} (\sim0$; middle), and strong 
correlations with a negative $r_{12}$ (bottom) for type 1 AGN (left) and 
type 2 AGN (right).}
}
\end{figure*}

For type 1 AGN, BH mass is calculated using the virial method, in which BH mass
($M_{\rm BH} \sim RV^2/G$) can be inferred from the size ($R$) and velocity 
dispersion ($V$) of the BLR underneath the central black hole. 
Thanks to the tight correlation between $R$ and AGN luminosity ($R-L$ relation),
one can estimate BH mass using the single epoch spectrum. 
However, in the virial mass estimates, there is an unknown scaling factor $f$, 
which is sensitive to geometry and kinematics of gas clouds in BLR. 
In general, $f$ is determined by assuming that host galaxies of AGN follow
the same \msig\ relation of normal galaxies \citep[e.g.,][]{onken_2004, woo_2013}. 
As a result, a single universal 
scaling factor is used to estimate BH mass, although it can depend on 
physical properties of AGN \citep[e.g.,][]{marconi_2008}. 
Therefore, in addition to the intrinsic 
scatter in the $R-L$ relation, the unknown scaling factor can naturally introduce 
an uncertainty ($\sim0.5$ dex) to BH mass derived from the virial method
\citep[e.g.,][]{park_2012}. 

 Line width (full width at half maximum; FWHM) and 
flux of broad emissions (\hb\ and \hal)
were taken from \citet{rakshit_2020}, whose values
are measured from the decomposition of SDSS spectra.
The broad emission lines were fitted with multiple-Gaussian models.
We adopted a calibration of BH mass estimator from \citet{ho_2015} 
based on \hb\ and 5100\AA\ AGN continuum luminosity \citep{kim_2021a}.
We primarily utilized \hal\ line widths and \hal\ luminosity 
as \hal\ is minimally influenced by the complex Fe II multiplets 
than \hb. For the conversions of line luminosity to 5100\AA\ AGN luminosity, 
and \hal\ line widths to \hb\ line widths, we employed the relations 
of \citet{greene_2005}.
For those without
reliable spectral measurements for \hal, we utilized \hb\ instead.  Overall, 
BH mass can be calculated from the virial method
for 2112 out of 2114 type 1. 

\begin{figure*}[t!]
\centering
\includegraphics[width=0.95\textwidth]{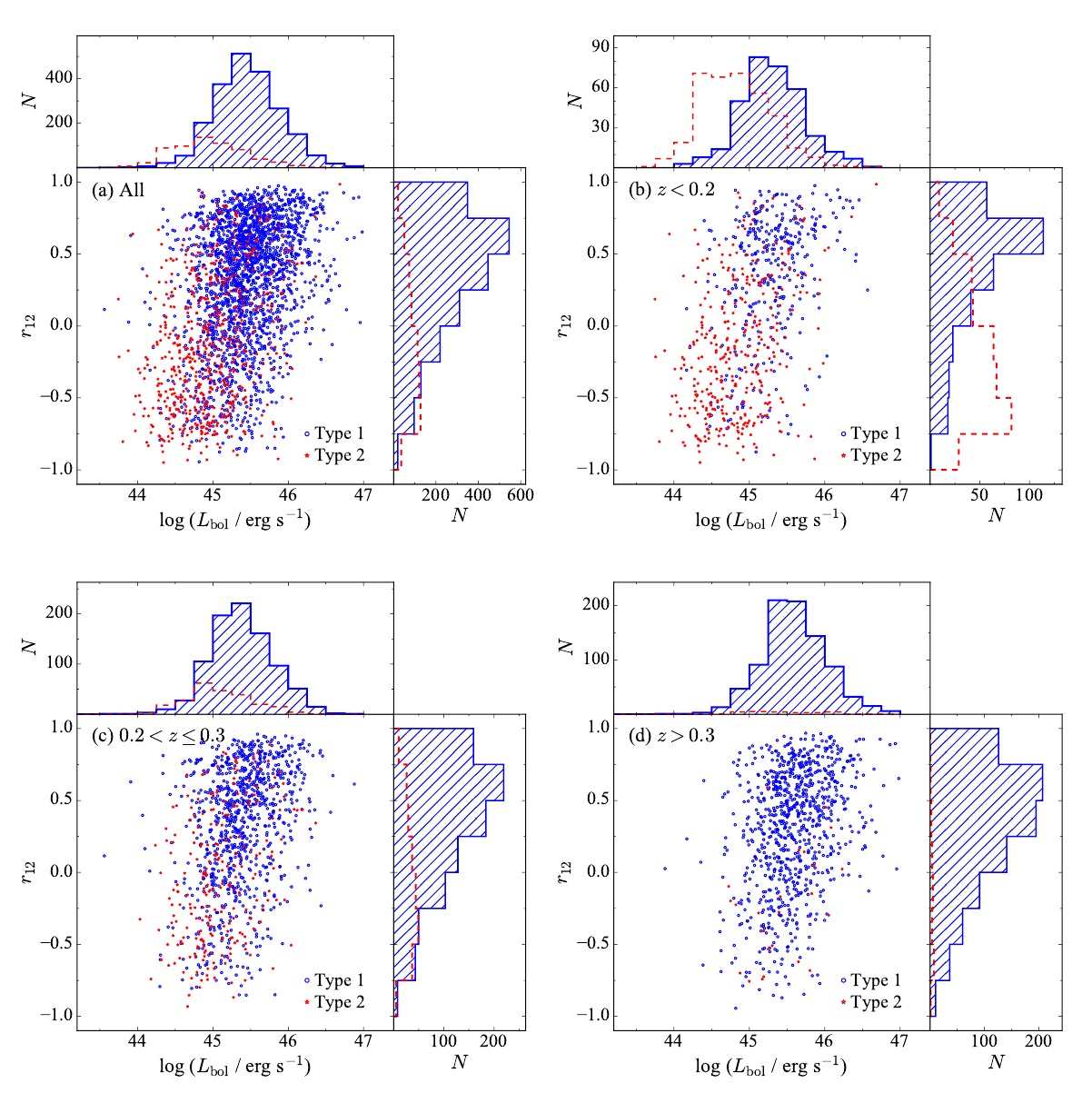}
\caption{
Dependence of $r_{12}$ on AGN bolometric luminosity 
for (a) the entire sample; (b) objects with $0.15 < z \leq 0.2$;
(c) objects with $0.2 < z \leq 0.3$; (d) objects with $0.3 < z \leq 0.4$. 
Type 1 (Type 2) objects are denote by blue stars and blue histograms 
(red circles and red histograms).   
}
\end{figure*}

\begin{figure*}[t!]
\centering
\includegraphics[width=0.95\textwidth]{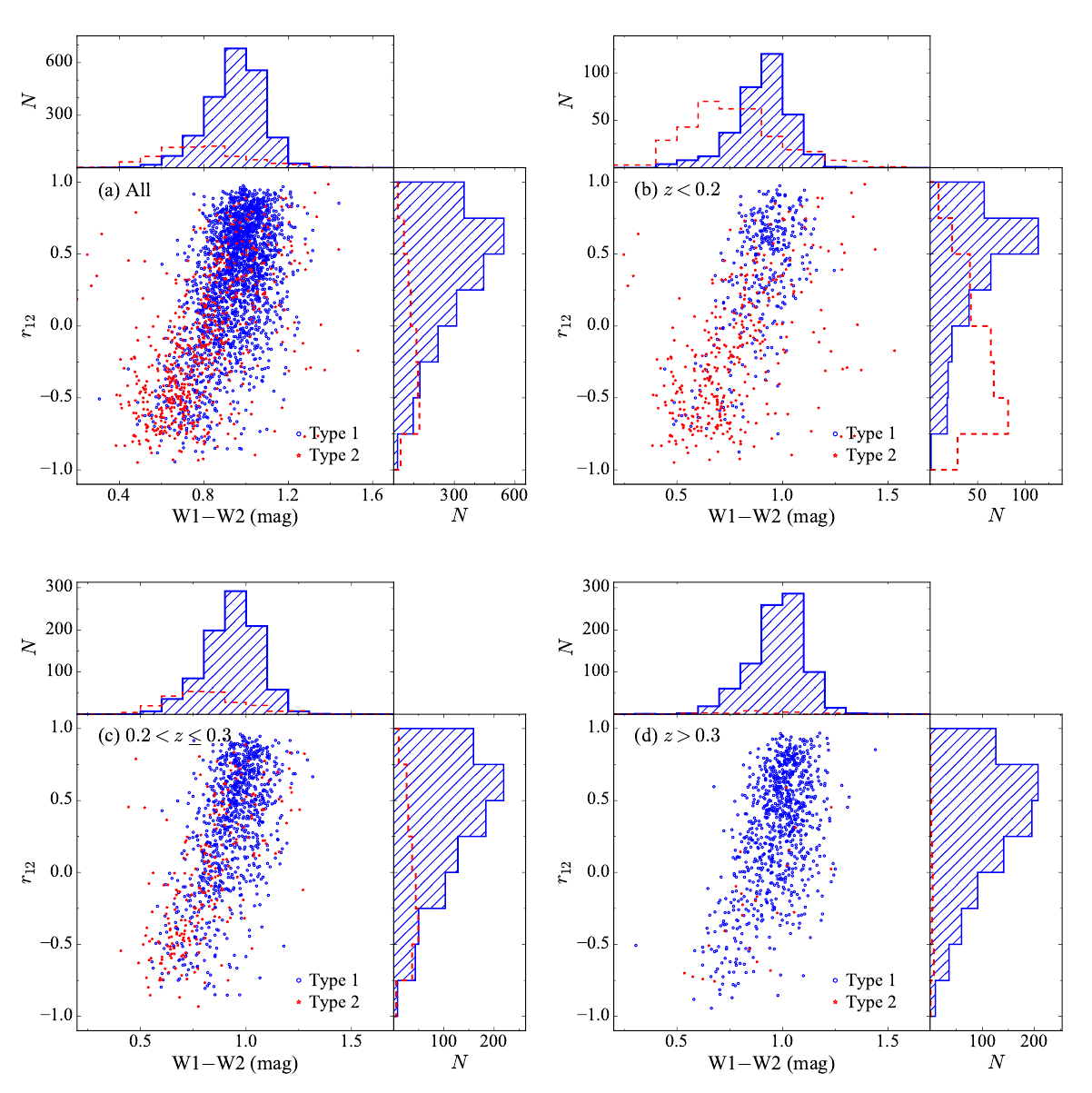}
\caption{
Dependence of $r_{12}$ on W1$-$W2       
for (a) the entire sample; (b) objects with $0.15 < z \leq 0.2$;
(c) objects with $0.2 < z \leq 0.3$; (d) objects with $0.3 < z \leq 0.4$.
The symbols are the same as in Figure 4.
}
\end{figure*}

\begin{figure*}[t!]
\centering
\includegraphics[width=0.95\textwidth]{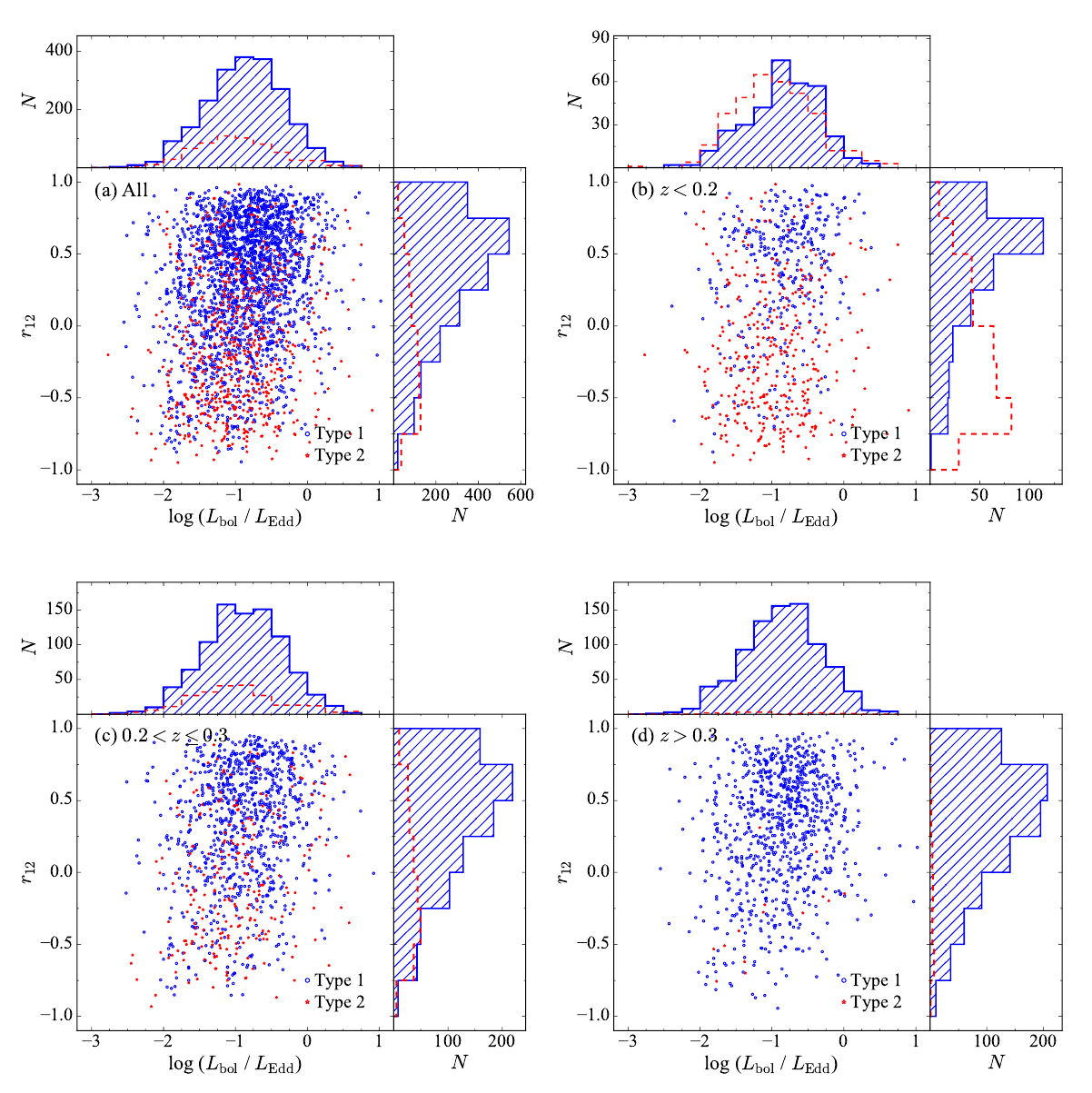}
\caption{
Dependence of $r_{12}$ on Eddington ratio       
for (a) the entire sample; (b) objects with $0.15 < z \leq 0.2$;
(c) objects with $0.2 < z \leq 0.3$; (d) objects with $0.3 < z \leq 0.4$.
The symbols are the same as in Figure 4.
}
\end{figure*}

\begin{figure*}[t!]
\centering
\includegraphics[width=0.95\textwidth]{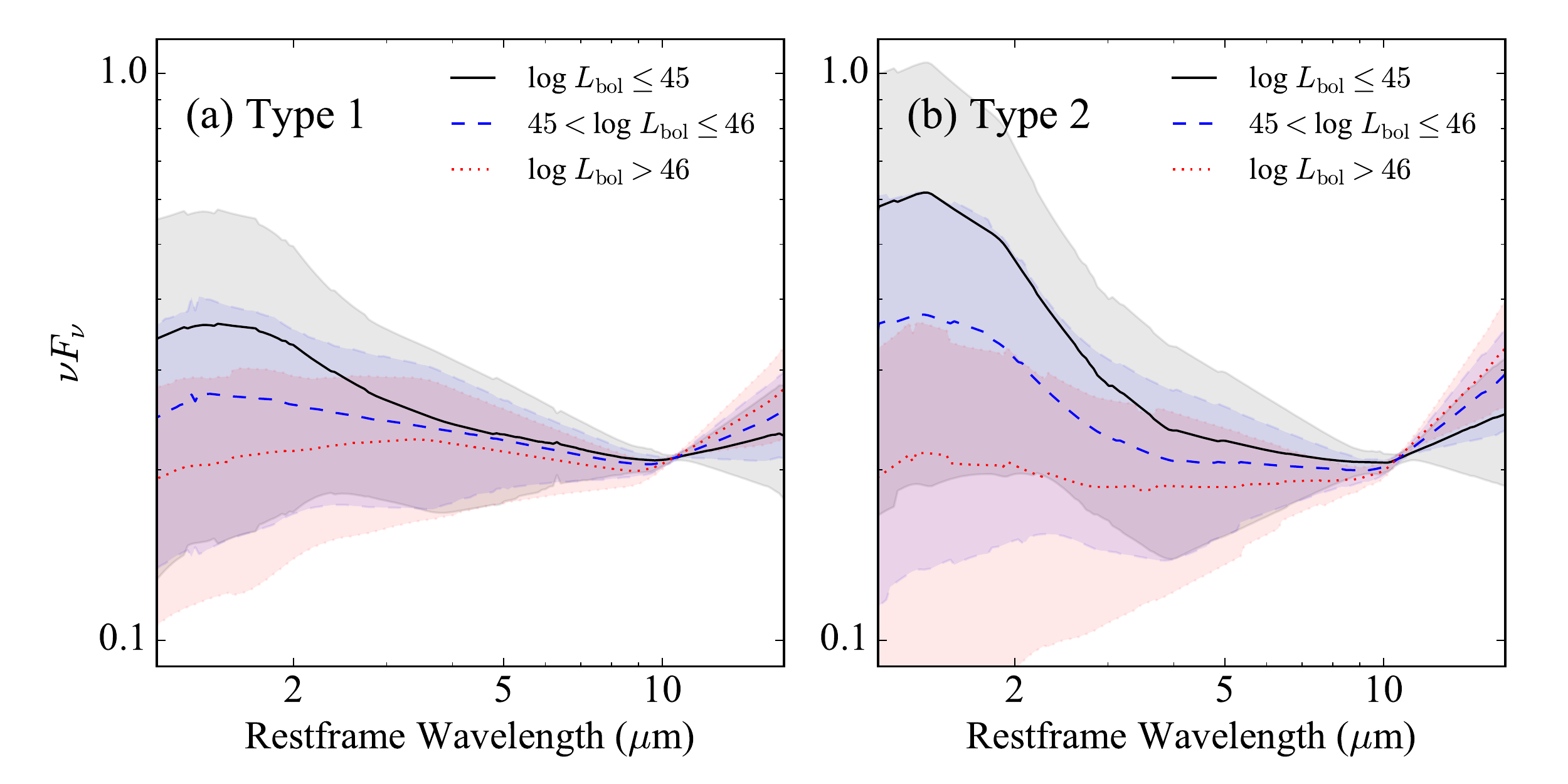}
\caption{
Composite spectra of (a) type 1 AGN and (b) type 2 AGN. In each panel, 
we divide the sample according to the bolometric luminosity: 
$\log ($\lbol$/$\lum$)$ $\le 45$ (black line),
$45 < \log ($\lbol$/$\lum$)$ $\le 45$ (blue dashed line),
$46 < \log ($\lbol$/$\lum$)$ (red dotted line). The shaded regions represent
$1\sigma$ dispersions of the composite spectra. 
}
\end{figure*}

For type 2 AGN, BH mass was computed from the total stellar mass of 
the host galaxy using the \mm\ relation. We adopted the \mm\ relation 
from \citet{greene_2020} for all morphological types of galaxies. 
Note that the uncertainty of this BH mass estimate ($\sim0.8$ dex) 
is significantly greater than that ($\sim 0.5$ dex) 
of the virial method due to the intrinsic scatter in the \mm\ relation 
\citep{greene_2020}. The stellar mass, derived from
the SED fit for the photometric data of {\it ugriz}, was taken from MPA-JHU 
catalog for 613 out of 623 type 2. Figure 3 shows the distributions of 
BH mass and Eddington ratio of the variable sources, in which there
is little difference between type 1 and type 2 AGN,   
contrary to the distributions of bolometric luminosity and redshift, 

\section{Result}

\subsection{The Connection between MIR Variability and AGN Properties}
In order to quantify the characteristics of MIR variability of an individual
AGN from the light curve, we utilized the Spearman correlation coefficient 
between W1 (mag) and W1$-$W2 color (hereafter $r_{12}$). 
If W1$-$W2 color decreases (i.e. bluer in color) with increasing AGN 
brightness in the light curve, $r_{12}$ is predicted to be positive (Fig. 4). 
However, if MIR color becomes redder as AGN brightness increases,
$r_{12}$ is predicted to be negative.
Surprisingly, $r_{12}$ is found to have a broad distribution from $-1$ to 1,
indicating the structure of the torus may be complicated and different 
between AGN possibly dependent upon AGN properties.  
To investigate the physical origin of the wide spread of $r_{12}$, we 
performed the comparison of $r_{12}$ with various physical properties. We found that 
$r_{12}$ is moderately correlated with W1$-$W2 color ($r_s=0.45$ and $0.59$ for 
type 1 and type 2, respectively), and mildly correlated with the bolometric 
luminosity ($r_s=0.28$ and 0.32 for type 1 and type 2, respectively; Figs. 5 
and 6). P-values are smaller than $0.001$ in both cases, implying that
the null hypothesis (no correlation) can be rejected.
However, $r_{12}$ appears to be weakly or not correlated with 
the Eddington ratio (Fig. 7 and Tab. 1). 

The rest-frame central wavelengths of W1 and W2 vary with the redshift and 
hence the light contamination from the host and accretion disk can also vary 
with the redshift. In addition, the distributions of bolometric 
luminosity and redshift is clearly different between type 1 and 
type 2 AGN. 
To minimize those biases, we divide our sample into three subsamples 
according to their redshift. Our finding that $r_{12}$ is
correlated with W1$-$W2 color and bolometric luminosity is the same 
regardless of the redshift (Figs. 5 and 6). At $0.2<z\le0.3$, where
the difference between type 1 and type 2 AGN in the distribution of 
bolometric luminosity is minimized, the trend is almost identical
for both AGN types. It may reveal that the overall trend may not be 
affected by the bias in the distribution of bolometric luminosity. 

Interestingly, the distributions of $r_{12}$ are distinguishable between 
type 1 and type 2, in the sense that the mean $r_{12}$ of type 2 
($\sim-0.13\pm 0.46$) is systematically smaller than that of type 1 
($\sim 0.33\pm 0.42$). This may be partly due to the fact that type 1
tends to have larger bolometric luminosity and redder W1$-$W2 color compared
to type 2. This may be due to the fact that SDSS spectroscopic data
were obtained for a flux-limited sample ($r <17.77$). Therefore, for type 2
objects, the sample is biased toward AGN with more luminous hosts, leading to 
bluer W1$-$W2 color. On the contrary, for type 1 sources, owing to the light 
contribution from the AGN itself in the optical band, can be relatively 
free from this bias. While AGN tend to have red color in W1$-$W2 
(i.e. W1$-$W2 $> 0.8$; \citealt{stern_2012}), a large fraction of type 2s 
appear to have W1-W2 less than 0.8, which again suggests the substantial 
contribution from the host in these objects. We will discuss this in more detail 
in \S{4.1}.  
However, the correlation coefficient between $r_{12}$
and W1$-$W2 color appeared to be systematically greater for type 2 than type 1
(0.59 vs. 0.45), possibly implying that the genuine torus structure or
the viewing angles of type 1 and type 2 AGN are not identical. Because, 
however, not only the AGN properties of the sample (e.g., bolometric 
luminosity) are 
variable between type 1 and type 2, but also the host contamination is 
non-negligible in MIR, indicating the need for those factors to be fully taken 
into account to investigate the physical origin of our finding (see \S{4.1})

\begin{figure}[h]
\centering
\includegraphics[width=0.45\textwidth]{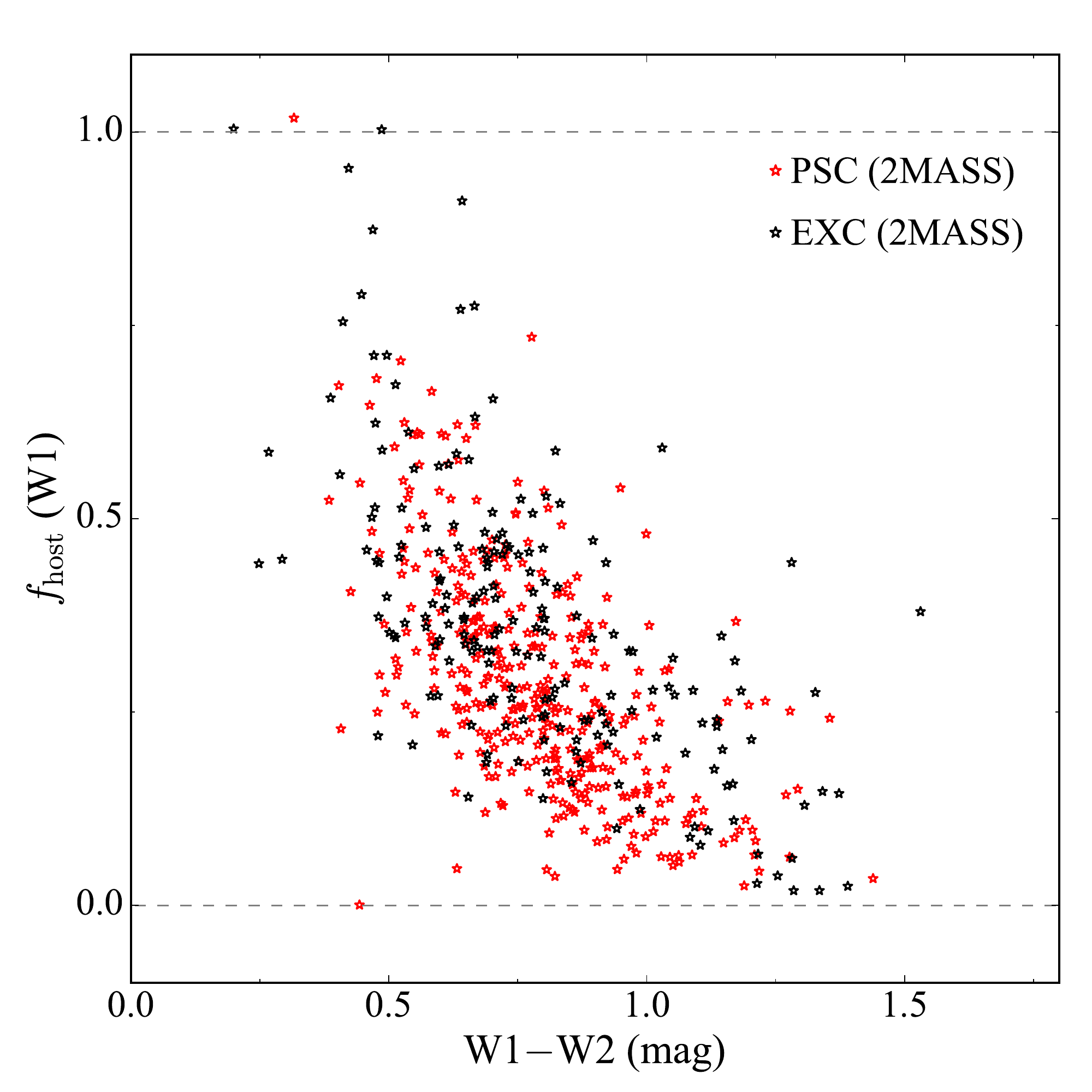}
\caption{
Light fraction of host galaxy at W1 band for type 2 AGN, inferred from the 
stellar mass calculated from the optical SED. Extended (EXC) and point sources 
(PSC) in 
2MASS images are denoted by red and black stars, respectively. 
}
\end{figure}

\subsection{Composite Spectra}
To investigate the origin of the observed dependence of $r_{12}$ on \lbol,
we constructed NIR/MIR composite spectra employing the photometric data from
2MASS ($JHK_s$; \citealt{skrutskie_2006}) and WISE (W1, W2, W3, and W4). 
We used a matching radius of 2\asec\ for the cross-correlation of WISE with 
2MASS. Note that the optical data is excluded because the light
contribution from the accretion disk is severe at short wavelenghts,which 
makes a direct comparison between two AGN types difficult.
Following the method from \citet{hickox_2017}, we included objects 
detected in all 7 bands (1866 type 1 and 559 type 2 AGN). 
For each object, the photometric data was converted to the rest-frame 
and fluxes in wavelengths ranging from $1-17\mu$m are calculated
by interpolating the photometric data in log-log space. The interpolated
spectrum of each object is normalized to integrated flux within $8-13\mu$m. 
The composite spectrum is established by averaging fluxes from all 
the sources in each spectral bin. To prevent the systematic bias owing to 
outliers, we exclude fluxes using iterations of $3\sigma$ clipping. 
We divide the sample into six subsamples based on AGN type and \lbol, to 
investigate how the SED shape depends on those parameters (Fig. 8).
Overall, 252, 785, and 829 type 1 objects (321, 171, and 67 type 2 objects) 
were employed to generate the composite spectra for low-luminosity 
($\log L_{\rm bol} \le 45$
\lum), moderate-luminosity($45 < \log L_{\rm bol} \le 46$
\lum), and high-luminosity objects ($\log L_{\rm bol} > 46$
\lum), respectively. In both types of AGN, more luminous AGN tend have 
redder MIR color compared to less luminous AGN. In addition, the signature of 
the host galaxy light in NIR appeared to be more outstanding in type 2 AGN. 
This is likely because type 2 AGN are systematically biased
toward having greater NIR emission from the host galaxy owing to the shallowness
of t`he 2MASS survey. It demonstrates the importance of  
quantifying the host contribution to understand the characteristics of the MIR 
variability.
\begin{figure*}[t!]
\centering
\includegraphics[width=0.95\textwidth]{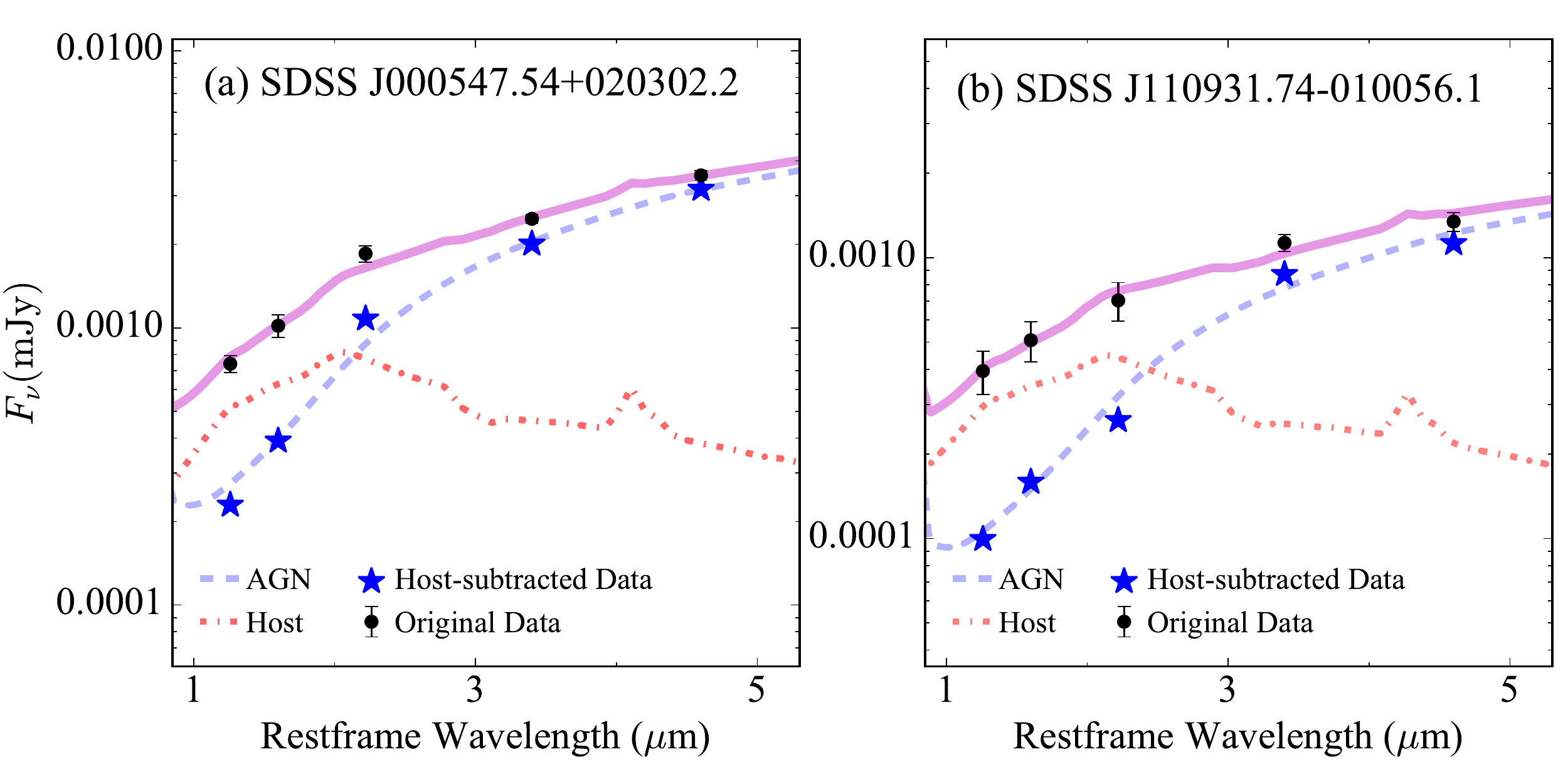}
\caption{
Examples of the SED fit for (a) type 1 and (b) type 2 AGN. Black circles 
represent the photometric data from 2MASS ($JHK_s$) and WISE (W1 and W2 bands). 
Blue dashed 
line, red dashed-dotted line, and magenta line represent SED models
for AGN, host galaxy, and the sum of two components, respectively. 
Blue stars represent the host-subtracted SED.
}
\end{figure*}

\section{Discussions}
\subsection{Host Contamination}
We found that $r_{12}$ is strongly correlated with W1$-$W2 color and 
bolometric luminosity. Objects with bluer MIR color or lower 
luminosity tend to have smaller $r_{12}$ than those with redder MIR color or 
higher luminosity. Taken at a face value, this may indicate that
the structure of the dusty torus is directly dependent upon 
the bolometric luminosity and subsequently the MIR color. 
However, this trend can be naturally explained if MIR light is dominated by 
the host component. In contrast to the AGN light, the stellar light is 
intrinsically blue in MIR and  does not vary on a time scale of years. 
As a result, if AGN outshines the host light in the low-luminosity
and host-dominated AGN, MIR becomes brighter and redder (i.e. $r_{12} < 0$).  
Therefore, it is crucial to robustly take the host contribution into 
account. Here, we utilized two different methods to estimate the relative host 
contribution in W1 and W2 band. First, to compute the predicted flux in W1 
predominantly from the stellar light, we utilized the stellar mass inferred 
from the SED fit of the optical broad band ({\it ugriz}) photometry
of type 2 AGN \citep{kauffmann_2003}. 
Note that the same method cannot be applied to type 1s because 
the light contribution from the accretion disk is significant.

The mass-to-light ratio of passive galaxies ($\sim 0.2$) at the W1 band 
was adopted from \citet{kettlety_2018}, as the MIR SED 
of the AGN host is established to be well fit with that of early-type galaxies 
\citep{hickox_2017}.
$k-$correction is calculated by presuming that MIR SED took a power
law following the recipe in \citet{kettlety_2018}. 
We adopted the average flux in the W1 band over the entire period of 
the WISE survey to compute the host contribution.
On average, the light fraction of the host galaxy 
in W1 band was $28\%$, implying that the host contribution is substantial
(Fig. 9).
This fraction dramatically increases as W1-W2 decreases, revealing 
the MIR color is closely correlated 
with the host contribution, which needs to be properly accounted for when 
calculating $r_{12}$. Note that due to the simplified assumptions on 
the galaxy template in this method, the result 
needs to be interpreted with caution.

Second, to robustly remove the light contamination from the host, we applied 
the SED fit to the broadband photometry of each object. For the SED fit, we 
utilized photometric data from 2MASS ($JHK_s$) and WISE (W1 and W2 bands). 
Because the SED fit with W3 and W4 bands often leads to unsatisfactory results,
those were excluded in this experiment. As W1 and W2 fluxes were measured 
through the PSF photometry, it can miss the flux from extended structures if 
present. Therefore, to keep the consistency between the photometric data, 
we only utilized the objects in the 2MASS point source catalog 
(PSC). 
We employed the template spectra for AGN and for an elliptical, 
spiral, and irregular galaxy from \citet{assef_2010}. 
A broad range of galaxy types provides better coverage of the diverse 
stellar population expected for the host galaxies of type 1 and type 2 AGN 
(e.g., \citealt{kim_2019,zhao_2019,zhao_2021}). During the fit, the extinction 
for the AGN template is considered by varying $A_V$.
From the SED fit, we computed the light contributions from the host in W1 and 
W2 bands, and subtract those from photometric data in every epochs 
(Fig. 10). Finally, we recalculate $r_{12}$ using the host-subtracted 
magnitudes in W1 and W2 bands.       

\begin{figure*}[t!]
\centering
\includegraphics[width=0.95\textwidth]{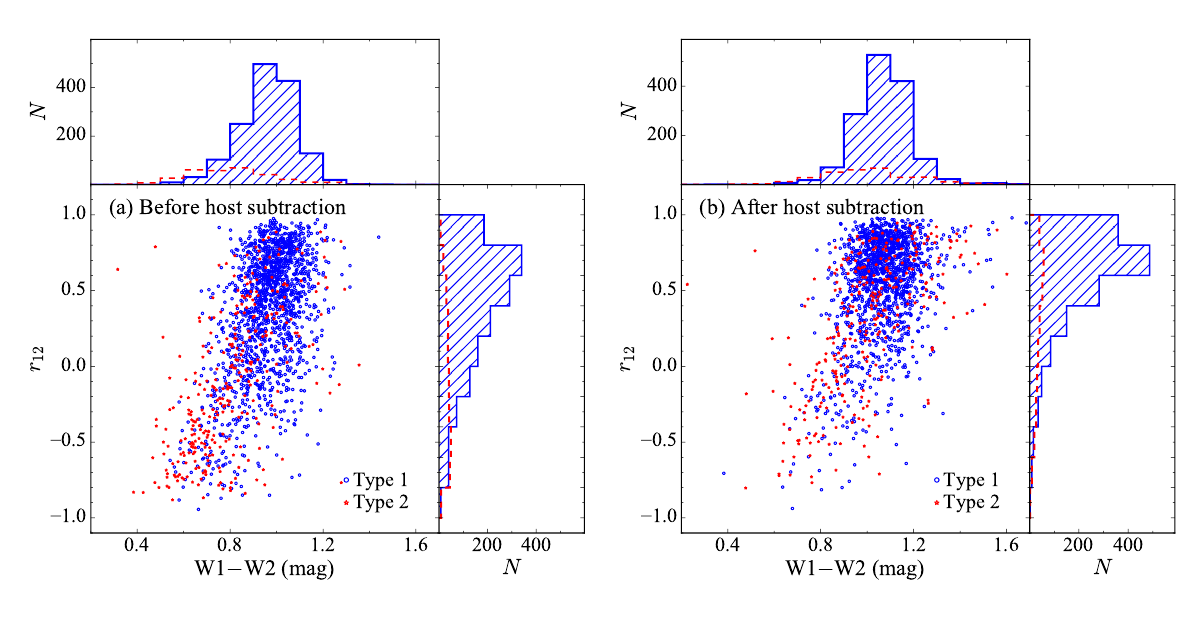}
\caption{
Correlations between $r_{12}$ and W1$-$W2 color (a) before and
(b) after the subtraction of host light. The symbols and lines are the same
as for Fig. 6.
}
\end{figure*}

\begin{figure*}[t!]
\centering
\includegraphics[width=0.95\textwidth]{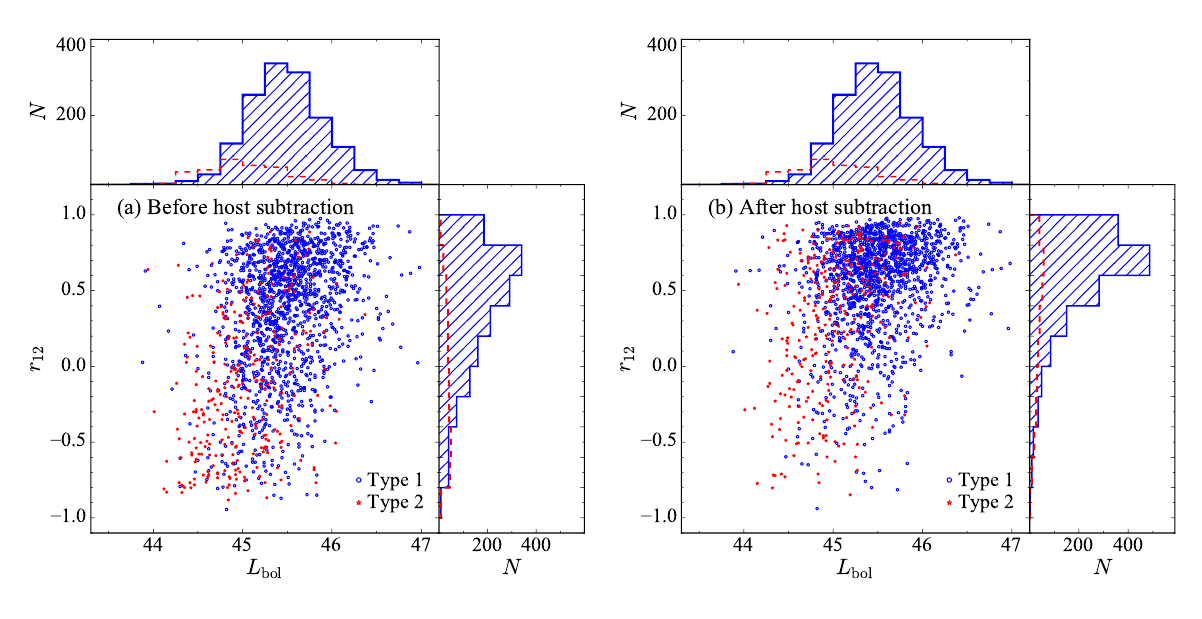}
\caption{
Correlations between $r_{12}$ and AGN bolometric luminosity (a) before and
(b) after the subtraction of host light. The symbols and lines are the same
as for Fig. 6.
}
\end{figure*}

Figure 11 (Figure 12) show the correlations between $r_{12}$ and W1$-$W2 
color (bolometric luminosity) before and after the subtraction of host light. 
The fractions of host light in W1 were, on average, {\bf 22 and 39}\% in 
type 1 and type 2 AGN, respectively. This is in broad agreement 
with the results from the stellar mass for type 2 AGN. 
However, the SED fit yielded 
a slightly larger host fraction in W1, possibly because both methods used 
templates from the different types of host galaxies.
%the sample for the SED fit only contained the point sources in 2MASS, 
%for which the host contamination 
%can be systematically smaller than that for extended sources. 
The host light subtraction naturally resulted in a marginal increase in 
W1$-$W2 color ($\sim0.1-0.2$) and a more substantial increase in
$r_{12}$ ($\sim0.19-0.38$ on average). Figure 11 clearly shows that,
after the correction, the majority of the sample have W1$-$W2 color larger 
than 0.8 mag, which is consistent with the previous criteria for the AGN 
selection (e.g., \citealt{stern_2012}).  
As a result, the level of dependence of 
$r_{12}$ on W1$-$W2 color became slightly lower only for type 1. 
Before the host correction,$r_s \sim 0.40$  for type 1 and 
$0.63$ for type 2, to be compared with $r_s \sim 0.24$ for type 1 and 
$0.64$ for type 2 after the correction. However, the general trend, in which
$r_{12}$ tends to decrease with decreasing W1$-$W, remains the same.    
This experiment reveals that the dependency on W1$-$W2 is not entirely
due to the host contamination, rather it is more likely connected with 
the intrinsic properties of the dusty torus. Similarly, the trend 
for the dependence of $r_{12}$ on AGN luminosity also still remains the same
after the correction for the host galaxy.
The correlation coefficients between $r_{12}$ and physical parameters of 
AGN after the correction for the host galaxy are listed in Table 1.

\begin{figure}[t!]
\centering
\includegraphics[width=0.45\textwidth]{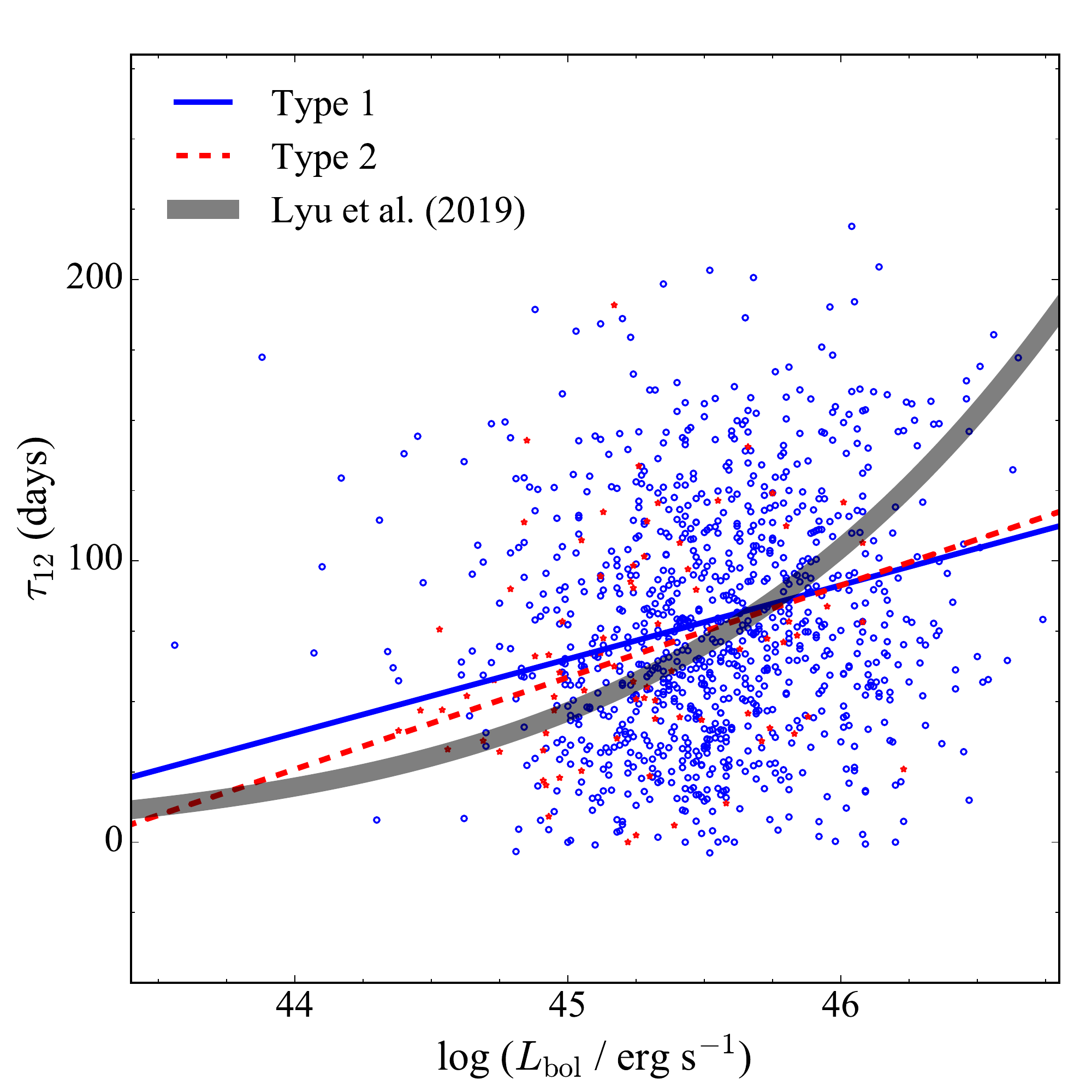}
\caption{
Distributions of time lag between W1 and W2 as a function of bolometric 
luminosity. Type 1 and Type 2 objects are denoted by open blue circles
and red stars, respectively. The fitting results of linear regression 
are shown as blue solid line and red dashed line for Type 1 and Type 2 
objects, respectively. The thick gray line represents the relation 
inferred from Lyu et al. (2019).
}
\end{figure}

\subsection{Physical Origin of the Wide Range of $r_{12}$}
The brightness and color in W1 and W2, can be determined from several physical 
parameters, such as extinction for the AGN ($A_V$), existence and 
covering 
factor of hot dust component, viewing angle, and host galaxy contamination 
(e.g., \citealt{nenkova_2008, stalevski_2016, hickox_2017, lyu_2017}). 
Our findings imply that the characteristics of variability 
in W1 and W2 are strongly dependent upon W1$-$W2 color and AGN luminosity.    
In \S{4.1}, we demonstrate that the host contamination cannot be a physical origin
of such trends. In addition, the viewing angle may not change 
dramatically in a single object. Therefore, we conclude that
the hot dust component or the obscuration may be the main driver of the variability
of W1$-$W2 color.

\subsubsection{What makes $r_{12}$ positive?}
The innermost radius of the torus is determined by the sublimation of the 
dust \citep{barvainis_1987, suganuma_2006}. Therefore it is well known that 
hotter dust component have a smaller sublimation 
radius than warmer dust component, indicating that dust emission from 
shorter wavelength originates from an inner part of the torus 
(\citealt{koshida_2014, lyu_2019}).
Therefore the reprocessed emission in W1 responded to the variation of the 
continuum from the accretion disk more quickly than that in W2. As a result,
W1$-$W2 color became blue (red) with increasing (decreasing) AGN luminosity,
yielding a positive $r_{12}$ unless there is no change in the structure 
of the torus. 
To confirm this hypothesis, we estimate the time lag between W1 and W2
based on the light curve employing interpolated cross-correlation functions
(ICCF; \citealt{peterson_1998, sun_2018}). We consider the centroid of the 
ICCF as the time lag. Errors on the 
time lags were estimated from Monte Carlo (MC) iterations \citep{gaskel_1987}. 
Figure 13 shows that the time lag is greater than zero, revealing 
that W1 emission radiates from the inner part of the torus compared to 
W2. We also find that the time lag is correlated with \lbol, although with a 
large scatter, which is in broad agreement with the results from previous 
studies (\citealt{lyu_2019}). The correlation appears to be very weak 
due to the sparse sampling ($\sim6$ months) of the WISE light curves.

\begin{figure}[t!]
\centering
\includegraphics[width=0.45\textwidth]{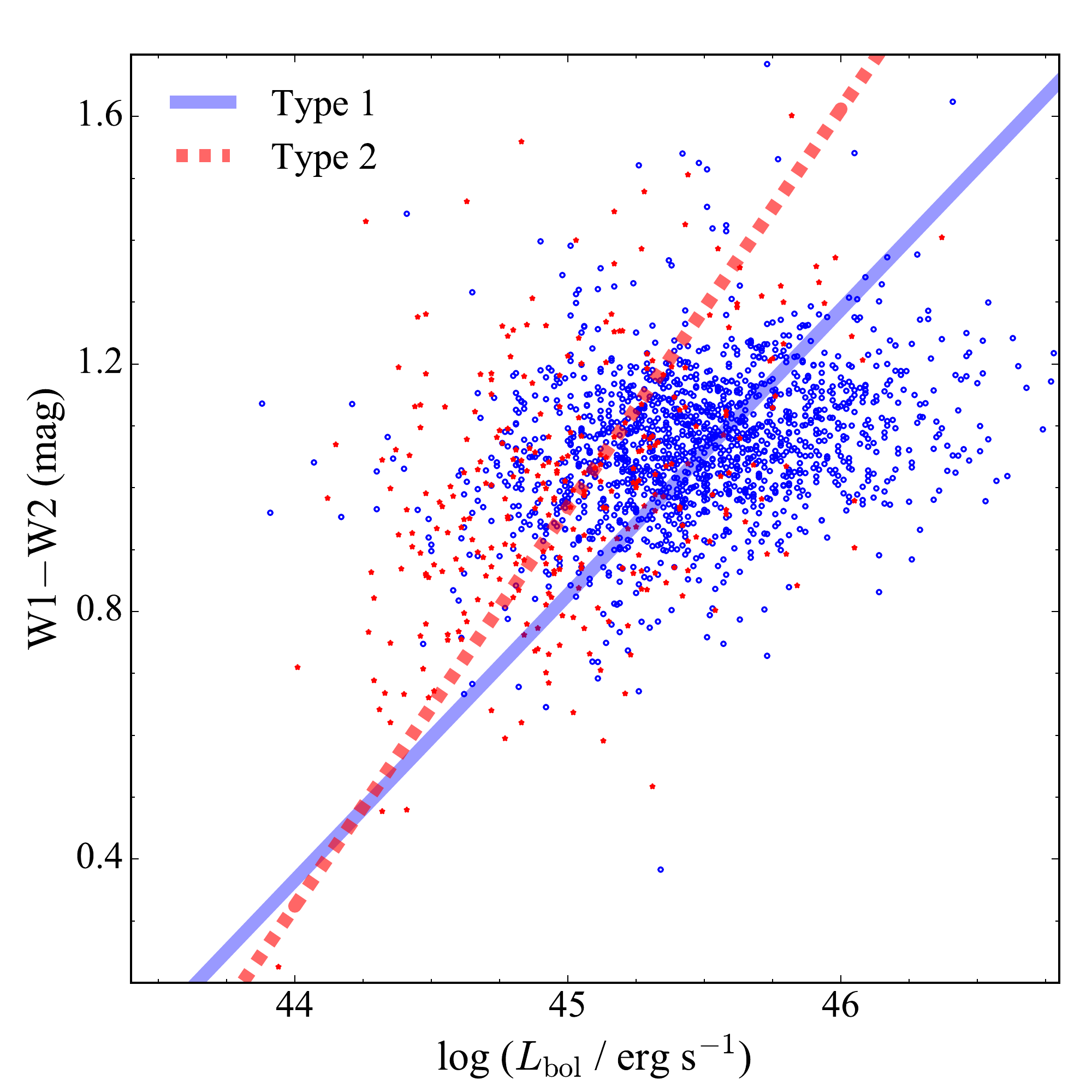}
\caption{
Relationship between W1$-$W2 color and AGN bolometric luminosity. The symbols
and lines are the same as for Fig. 13.  
W1$-$W2 color is corrected for the host galaxy contribution.
}
\end{figure}
\subsubsection{Viewing Angle and Obscuration}
The level of dependence of W1$-$W2 color and \lbol\ on $r_{12}$ is greater for
type 2 than for type 1 AGN irrespective of taking the host contamination into
account. This is partially owing to the fact that type 1 AGN lack objects with
$r_{12} < 0$. Therefore it may indicate either the contribution from the
accretion disk is not negligible even in W1 or the genuine structures of
the dusty torus between type 1 and type 2 AGN may vary. The viewing angle
and obscuration may also play an important role in shaping the NIR/MIR SED, 
in the sense that the smaller viewing angle and lower extinction
can result in the bluer color in W1 and W2 (e.g., \citealt{nenkova_2008,
stalevski_2016}). In this light, a marginal difference in $r_{12}$
distributions between type 1 and type 2 AGN can be described by those
parameters. 

According to the AGN unification model, type 1 AGN is
expected to have a smaller viewing angle and less obscuration than type 2 AGN.
As a result, if the extinction and viewing angle is predominantly responsible for 
W1$-$W2 color, type 1 AGN ought to have a bluer W1$-$W2 color than type 2 AGN,
which is inconsistent with our finding that type 2 AGN have
a slightly bluer W1$-$W2 color than type 1 AGN. In addition, Figure 14
demonstrates that higher luminosity AGN tend to have redder color, which 
is also contrary to the fact that the obscured AGN fraction decreases with 
the AGN luminosity (e.g., \citealt{maiolino_2007, lusso_2013}), if the 
red color is due to the high obscuration. It implies that neither the 
obscuration nor viewing angle play the primary role in determining W1$-$W2 
color.

\subsubsection{Covering Factor of Hot Dust Component}
It is known that the NIR/MIR SED of AGN is well fit with two components:
hot graphite dust component peaked around $2-3\mu$m (\citealt{richards_2006,
netzer_2007, mor_2012}) and warm silicate-type dust component
(e.g., \citealt{nenkova_2008}). Furthermore, the brightness in W1 and W2 are
dominated by the emission from the hot dust component \citep{mor_2011}. 
Therefore the variability and SED shape are likely sensitive to 
the physical properties of the hot component. 
\citet{mor_2011} demonstrated that the covering factor of the hot dust
component is anti-correlated with the AGN bolometric luminosity (\lbol). 
This is consistent with our finding that less luminous AGN have bluer
W1$-$W2 owing to the greater contribution from the hot dust component 
(Fig. 13). This can be also confirmed from the composite spectra, 
in the sense that low-luminosity AGN tend to have bluer color in $1-5\mu$m.
As a sanity check, we compared the AGN bolometric luminosity and the 
W1 luminosity ($L_{\rm W1}$), which were calculated in the rest-frame. 
$k-$correction were computed using the observed W1$-$W2 color. We found 
that the slopes between \lbol\ and $L_{\rm W1}$ ($\sim 0.76$ and $0.63$ for
type 1 and type 2, respectively) were smaller than 1 before taking 
account of the host contribution. However, after the subtraction
of the host light, the slopes became 0.92 and 0.91 for type 1 and 
type 2, respectively (Fig. 15). This finding
is in broad agreement with previous observational studies \citep{lusso_2013}, 
indicating that the covering factor of the hot dust decreases as 
AGN luminosity increases.  
Furthermore, there is a marginal discrepancy between type 1 and type 2 AGN
in this correlation. This result likely supports the scenario of the clumpy 
torus in which the hot component is less obscured than in the smooth torus 
model.

\begin{figure}[tbp]
\centering
\includegraphics[width=0.45\textwidth]{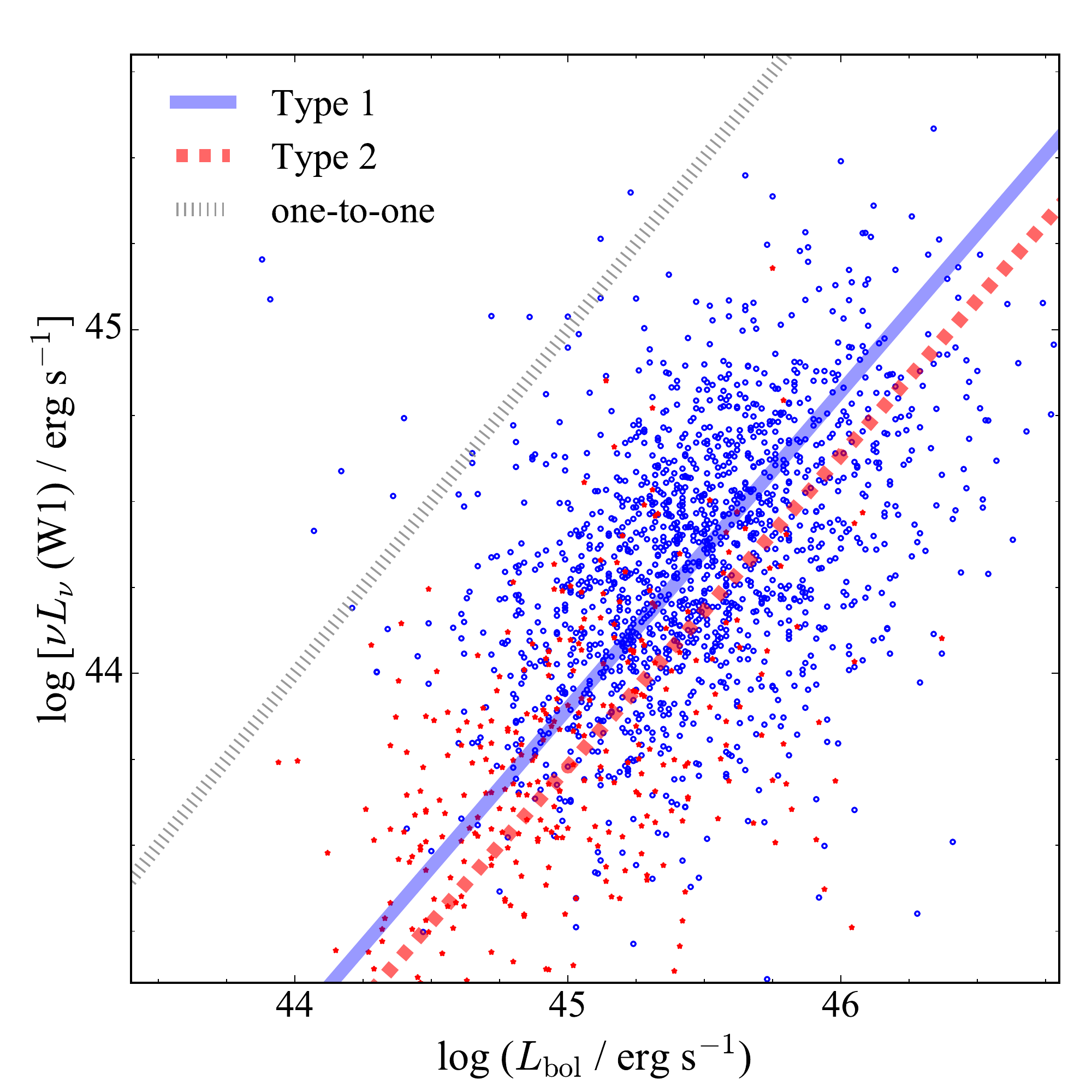}
\caption{
Relationship between W1 luminosity and AGN bolometric luminosity. The symbols
and lines are the same as for Fig. 13. The grey dotted line denotes one-to-one
correspondence. W1 luminosity is corrected for the host contribution.
}
\end{figure}

\begin{figure}[tbp]
\centering
\includegraphics[width=0.45\textwidth]{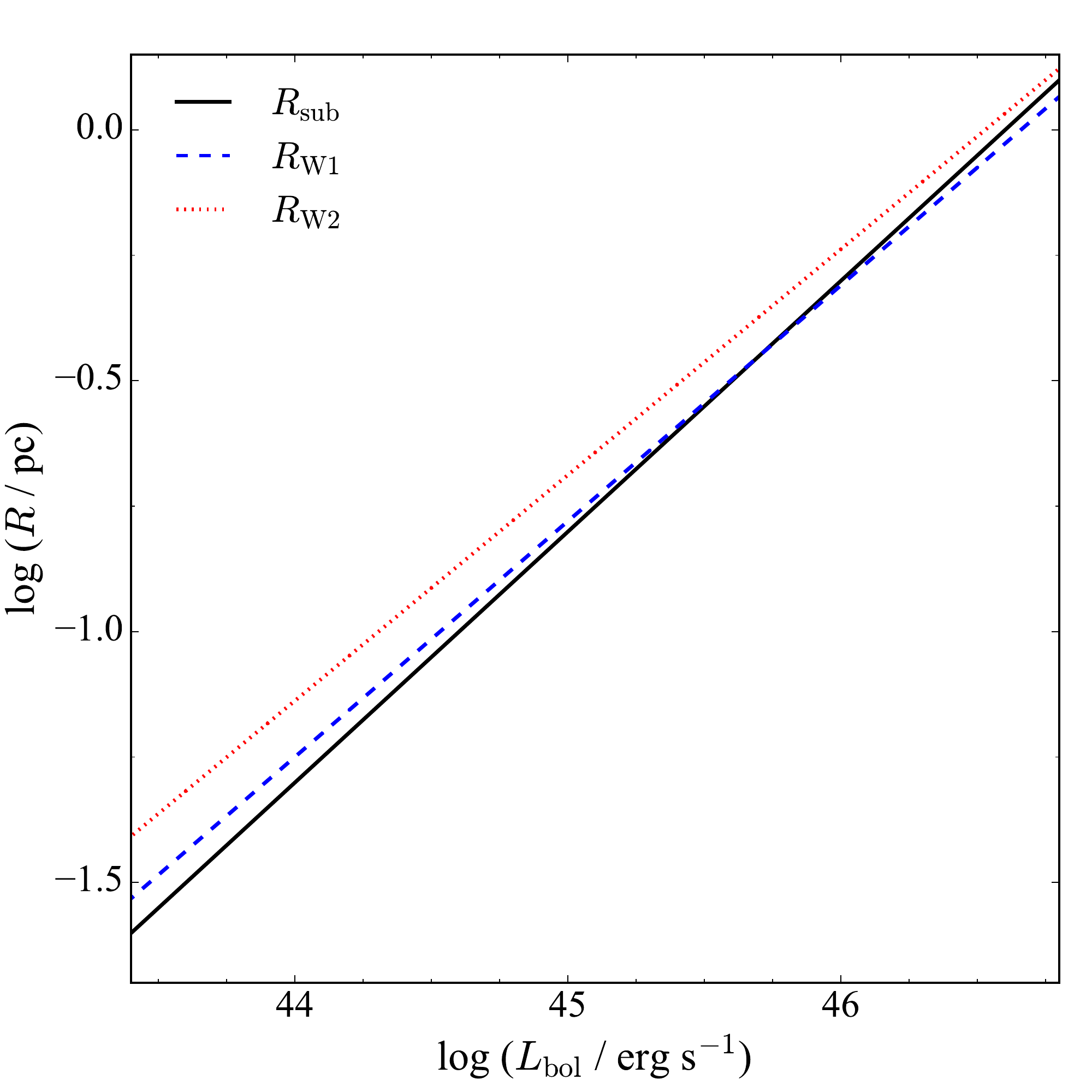}
\caption{
Relationship between AGN bolometric luminosity and various radii of the torus. 
Black solid line denotes the sublimation radius of the hot dust component. 
Blue dashed and red dotted lines represent the radii of the torus measured from
the reverberation mapping method in W1 and W2 bands, respectively 
\citep{lyu_2019}.  
}
\end{figure}

In addition, the sublimation radius ($R_{\rm sub}$) of the hot dust component 
is correlated with \lbol, given by :
\begin{equation}
R_{\rm sub} \approx 0.5 \left( \frac{L_{\rm bol}}{10^{46} {\rm erg s^{-1}}}
 \right)^{0.5}
\left( \frac{1800K}{T_{\rm sub}} \right)^{2.8} {\rm pc}, 
\end{equation}
where $T_{\rm sub}$ denotes a sublimation temperature of pure-graphite grains \citep{mor_2012}. 
Therefore, low-luminosity AGN tends to have a smaller innermost radius 
compared to high-luminosity AGN. One of the most interesting finding in our 
study is that, for low-luminosity AGN with blue W1$-$W2, W1$-$W2 color becomes 
redder with increasing AGN luminosity in the light curve (i.e. $r_{12} < 0$). 
This may indicate that the hot dust at the innermost radius can be easily 
evaporated by the enhanced emission from the accretion disk. 
Interestingly, \citet{yang_2018} reported that changing-look (CL) AGN 
also tend to have $r_{12} < 0$. This may imply that CL AGN event occurs
due to the change in the accretion rate.

\citet{lyu_2019} evaluated the radii ($R_{\rm W1}$ and $R_{\rm W2}$) of the 
torus inferred from the time-lag between the optical UV/optical emission and 
reprocessed emission in W1 and W2 band, respectively (Equations (15) and (16) 
of \citealt{lyu_2019}). They claimed that the torus size is strongly correlated 
with the AGN bolometric luminosity (see also 
\citealt{koshida_2014, glass_2004}). Figure 16 shows that the comparison 
between the sublimation radius and the torus radii. $R_{\rm sub}$ 
is almost identical to $R_{\rm W1}$ and significantly smaller 
than $R_{\rm W2}$, implying that dust material emitting in W1 band can be 
vulnerable to the sublimation. It confirms our argument is relevant.  
   
However, for the high-luminosity AGN with red W1$-$W2, the covering 
factor of the hot dust component can be reduced. Therefore, the NIR/MIR 
SED is unlikely sensitive to the evaporation of the hot dust and more likely 
determined by the warm dust component. 
In Figure 16, at $L_{\rm bol} \geq 10^{45.5}$\lum, $R_{\rm W1}$ becomes 
smaller than $R_{\rm sub}$, revealing that the hot graphite dust may be 
evaporated already.

Finally, polar dust has been found in high-resolution MIR images
of nearby type 1 AGNs (e.g., \citealt{honig_2013, stalevski_2017}). 
In addition, the light contribution from the polar dust may be 
substantial in the MIR SED (e.g., \citealt{lyu_2018, yang_2020, toba_2021}). 
Therefore, one might suspect whether the polar dust can be responsible for 
the trend found in this study. However, the location of the polar dust 
(i.e. close to sublimation radius or located in the host galaxies) is still 
under debate (e.g., \citealt{lyu_2018, yang_2020}). Moreover, the 
physical properties of the polar dust in type 2 are unknown. Therefore, 
investigating the contribution from the polar dust is beyond the scope of our 
study.

\section{Summary}
Using the light curves in W1 and W2 bands generated from WISE multi-epoch 
data, we found that W1$-$W2 color varies in respond to the light variation of 
AGN.
This indicates that the structure of the dusty torus is sensitive to AGN
luminosity. Furthermore, the trend of the color variation greatly
depends on the MIR color and AGN bolometric luminosity, in the sense
that low-luminosity (high-luminosity) AGN with blue (red) W1$-$W2 color
tend to turn redder (bluer) in W1$-$W2 color with increasing IR luminosity.
We found such trend is still valid, 
even after taking the host contribution in NIR/MIR into account, 
suggesting the host light is not the predominant 
reason for the color variation. We argue that the anti-correlation between 
the covering factor of the hot dust component and AGN luminosity can be the 
major reason for our findings, which is consistent with previous studies.
Our results demonstrate that the color variation in NIR/MIR inferred 
from the multi-epoch data can be served as a powerful tool to investigate 
the detailed structure of the torus. For example, SPHEREx mission will obtain 
multi-epoch spectral images in $\sim 200$ deg$^2$ of deep fields with a 
two-yr lifetime mission \citep{dore_2018, kim_2021b}. The multi-epoch spectral 
data covering $0.75-5.00\mu$m will provide a unique opportunity to address 
this subject in depth.

\acknowledgments
We are grateful for valuable comments from an anonymous reviewer, 
which helped to improve the paper.
This publication makes use of data products from the Wide-field Infrared Survey Explorer, which is a joint project of the University of California, Los Angeles, and the Jet Propulsion Laboratory/California Institute of Technology, funded by the National Aeronautics and Space Administration.
LCH was supported by the National Key R\&D Program of China (2016YFA0400702) 
and the  National Science Foundation of China (11721303, 11991052, 
12011540375). This work was supported by a National Research Foundation of 
Korea (NRF) grant (No.\ 2020R1A2C4001753) funded by the Korean government 
(MSIT) and under the framework of international cooperation program managed 
by the National Research Foundation of Korea (NRF-2020K2A9A2A06026245).

\begin{deluxetable*}{clcrccrccrccr}
\tablecolumns{13}
\tablenum{1}
\tablewidth{0pc}
\tablecaption{Spearman correlation coefficient ($r_s$) between
$r_{12}$ and various physical parameters}
\tablehead{
\colhead{\h} &
\colhead{\h} &
\multicolumn{2}{c}{Type 1} &
\colhead{\h} &
\multicolumn{2}{c}{Type 2} &
\colhead{\h} &
\multicolumn{2}{c}{Type 1 (Cor.)} &
\colhead{\h} &
\multicolumn{2}{c}{Type 2 (Cor.)} \\
\cline{3-4}
\cline{6-7}
\cline{9-10}
\cline{12-13}
\colhead{\h Parameter} &
\colhead{\h Sample} &
\colhead{\h $r_s$} &
\colhead{\h $p-$value} &
\colhead{\h} &
\colhead{\h $r_s$} &
\colhead{\h $p-$value} &
\colhead{\h} &
\colhead{\h $r_s$} &
\colhead{\h $p-$value} &
\colhead{\h} &
\colhead{\h $r_s$} &
\colhead{\h $p-$value} \\
\colhead{\h (1)} &
\colhead{\h (2)} &
\colhead{\h (3)} &
\colhead{\h (4)} &
\colhead{\h} &
\colhead{\h (5)} &
\colhead{\h (6)} &
\colhead{\h} &
\colhead{\h (7)} &
\colhead{\h (8)} &
\colhead{\h} &
\colhead{\h (9)} &
\colhead{\h (10)} 
}
\startdata
W1$-$W2 &\h All  &\h 0.45&\h$<0.001$&\h &\h 0.59&\h$<0.001$&\h &\h 0.24&\h$<0.001$&\h &\h 0.64&\h$<0.001$\\
\h &\h $0.15<z\leq0.2$   &\h 0.48&\h$<0.001$&\h &\h 0.53&\h$<0.001$&\h &\h 0.11&\h0.135&\h &\h 0.63&\h$<0.001$\\
\h &\h $0.2<z\leq0.3$  &\h 0.52&\h$<0.001$&\h &\h 0.68&\h$<0.001$&\h &\h 0.30&\h$<0.001$&\h &\h 0.67&\h$<0.001$\\
\h &\h $0.3<z\leq0.4$  &\h 0.42&\h$<0.001$&\h &\h 0.67&\h$<0.001$&\h &\h 0.22&\h$<0.001$&\h &\h 0.36&\h0.427\\
\hline
log $L_{\rm bol}$ &\h All  &\h 0.28&\h$<0.001$&\h &\h 0.32&\h$<0.001$&\h &\h 0.19&\h$<0.001$&\h &\h 0.32&\h$<0.001$\\
\h &\h $0.15<z\leq0.2$   &\h 0.31&\h$<0.001$&\h &\h 0.30&\h$<0.001$&\h &\h 0.22&\h0.003&\h &\h 0.39&\h$<0.001$\\
\h &\h $0.2<z\leq0.3$  &\h 0.32&\h$<0.001$&\h &\h 0.32&\h$<0.001$&\h &\h 0.29&\h$<0.001$&\h &\h 0.27&\h0.002\\
\h &\h $0.3<z\leq0.4$  &\h 0.30&\h$<0.001$&\h &\h 0.43&\h0.034&\h &\h 0.26&\h$<0.001$&\h &\h 0.72&\h0.068\\
\hline
$\log (L_{\rm bol}/L_{\rm Edd})$ &\h All  &\h 0.20&\h$<0.001$&\h &\h 0.12&\h0.002&\h &\h 0.13&\h$<0.001$&\h &\h 0.25&\h$<0.001$\\
\h &\h $0.15<z\leq0.2$   &\h 0.25&\h$<0.001$&\h &\h 0.03&\h0.552&\h &\h 0.10&\h0.166&\h &\h 0.23&\h0.002\\
\h &\h $0.2<z\leq0.3$  &\h 0.19&\h$<0.001$&\h &\h 0.27&\h$<0.001$&\h &\h 0.15&\h$<0.001$&\h &\h 0.26&\h0.003\\
\h &\h $0.3<z\leq0.4$  &\h 0.19&\h$<0.001$&\h &\h 0.02&\h0.943&\h &\h 0.11&\h0.009&\h &\h 0.65&\h0.111\\
\enddata
\tablecomments{
Col. (1): Physical parameter. 
Col. (2): Sample.
Col. (3): Spearman correlation coefficient for type 1 AGN.
Col. (4): Probability value that the correlation occurs by chance for type 1 AGN.
Col. (5): Spearman correlation coefficient for type 2 AGN.
Col. (6): Probability value that the correlation occurs by chance for type 2 AGN.
Col. (7): Spearman correlation coefficient for type 1 AGN after the correction for the host contribution. 
Col. (8): Probability value that the correlation occurs by chance for type 1 AGN after the correction for the host 
contribution.
Col. (9): Spearman correlation coefficient for type 2 AGN after the correction for the host contribution.
Col. (10): Probability value that the correlation occurs by chance for type 2 AGN after the correction for the host 
contribution.
}
\end{deluxetable*}

\bibliography{torus}

\end{document}